# Enhancement of $J_C$-B Properties in $MoSi_2$ doped $MgB_2$ tapes

Xianping Zhang[1,3], Yanwei Ma[1],[*], Zhaoshun Gao[1,3], Zhengguang Yu[1], G. Nishijima[2], K. Watanabe[2]

[1] Applied Superconductivity Lab., Institute of Electrical Engineering, Chinese Academy of Sciences, P. O. Box 2703, Beijing 100080, China

[2] High Field Laboratory for Superconducting Materials, Institute for Materials Research, Tohoku University, Sendai 980-8577, Japan

[3] Graduate University of Chinese Academy of Sciences, Beijing 100088, China

E-mail: ywma@mail.iee.ac.cn

**Abstract:**

$MoSi_2$ doped $MgB_2$ tapes with different doping levels were prepared through the *in-situ* powder-in-tube method using Fe as the sheath material. Effect of $MoSi_2$ doping on the $MgB_2$/Fe tapes was investigated. It is found that the highest $J_C$ value was achieved in the 2.5 at.% doped samples, more than a factor of 4 higher compared to the undoped tapes at 4.2 K, 10 T, then further increasing the doping ratio caused a reduction of $J_C$. Moreover, all doped tapes exhibited improved magnetic field dependence of Jc. The enhancement of $J_C$-B properties in $MoSi_2$ doped $MgB_2$ tapes is attributed to good grain linkage and the introduction of effective flux pining centers with the doping.

---

[*] Author to whom correspondence should be addressed

## 1. Introduction

The critical current density $J_C$ in $MgB_2$ has been a central topic for extensive research efforts since superconductivity in this metal compound was discovered [1]. $MgB_2$ has high superconducting transition temperature ($T_C$) and does not have weak-link problem at grain boundaries. The material cost of $MgB_2$ is lower compared to other superconducting materials such as Nb-based superconductors and HTS (High $T_C$ superconductivity) cuprates. Therefore, $MgB_2$ is a promising candidate for engineering applications at temperatures around 20 K. However, $J_C$ of $MgB_2$ decreases rapidly under magnetic fields compared to those for the Nb-based superconductors. It is generally agreed that the improvement of $J_C$ can be achieved in two ways: improvement of the grain connectivity and introduction of the effective pining centers [2]. In contrast to HTS, $MgB_2$ has a relatively large coherence length, and this means that the fluxiods can be pinned by relatively larger particles and precipitates. Accordingly, chemical doping seems to be a very promising way to increase the flux pinning and $B_{irr}$ in $MgB_2$ [3]. Many works have been done on the chemical doping/additive to $MgB_2$ [4–11]. Of all these materials, various silicides are very effective on the enhancement of $J_C$ in $MgB_2$ at high field regime [12, 13, 14]. In this work, a series of $MoSi_2$ doped $MgB_2$ tapes were prepared using an *in-situ* powder-in-tube (PIT) method. The difference in microstructure and critical current properties induced by various doping level is discussed.

## 2. Experimental details

$MgB_2$ tapes were prepared by the *in-situ* PIT method. The sheath materials chosen for this experiment were commercially available pure Fe. Mg (325 mesh, 99.8%), B (2-5 µm amorphous, 99.99%), and $MoSi_2$ (2-5 µm, 99%) powders were used as the starting material. Mg, B powders were mixed with the nominal composition of 2:1, the $MoSi_2$ doping levels were 2.5, 5, 10, 15, 20 at.%, respectively. After milling for 1 h, the mixed powder was tightly packed into Fe tubes of 8 mm outside diameter and 1.5 mm wall thickness. The composite tubes were subsequently swaged into rods of 5 mm in diameter, and then the rods were drawn into round wires about 1.5 mm in diameter. Subsequently, the wires were rolled to tapes. The final size of the tapes was 3.2 mm in width and 0.5 mm in thickness. Undoped tapes were similarly fabricated for comparative study. These tapes were sintered in flowing high purity Ar at 650°C for 1 h, followed by a furnace cooling to room temperature. To the study effects of sintering temperature on $MoSi_2$ doped $MgB_2$ tapes, samples heated at 750°C for 1 h were also made.

The phase composition and microstructure of the samples were investigated using X-ray diffraction (XRD), energy dispersive x-ray analysis (EDX) and scanning electron microscope (SEM). For SEM/EDX and XRD analysis, rectangular samples were cut from the tapes, and sheath materials were removed. DC magnetization measurement was performed with a superconducting quantum interference device (SQUID) magnetometer, using small pieces of an $MgB_2$ layer obtained by removing the sheath material of the samples. The $T_C$ was defined as the onset temperature at which a diamagnetic signal was observed. The transport $I_C$ at 4.2 K and its magnetic field dependence were evaluated at the High field laboratory for Superconducting Materials (HFLSM) at Sendai, by a standard four-probe technique, with a criterion of 1 μV cm$^{-1}$. Current leads and voltage taps were directly soldered to the sheath materials of the tapes. A magnetic field was applied parallel to the tape surface. The critical current density $J_C$ was obtained by dividing $I_C$ by the cross-sectional area of the $MgB_2$ core.

## 3. Results and discussion

Figure 1 exhibits XRD patterns of $MoSi_2$ doped and undoped samples. The XRD pattern for the undoped sample is consistent with the published indices of $MgB_2$ with a trace amount of MgO. For the 2.5 at.% doped samples, $MoSi_2$ phase is clearly seen as one of main impurities along with MgO. $MoSi_2$ phase can be identified in all the doped samples, and its diffraction peaks are increased with the increase of doping level. There is no observable shift in the peaks of the XRD patterns. This indicates that there were no changes in the lattice parameters between the undoped and $MoSi_2$ doped samples [15]. It is clear that $MgB_2$ is inert with respect to $MoSi_2$ at 650 . The similar result was also found for $WSi_2$-doped tapes, as reported by Ma et al [12]. On the contrary, the addition of $ZrSi_2$ in $MgB_2$ tapes resulted in the formation of $Zr_3Si_2$ and $Mg_2Si$, no diffraction peaks of $ZrSi_2$ phase was observed, suggesting that there were reaction between $MgB_2$ and $ZrSi_2$ [12]. Since Mo and W elements are in the same element group (VIb), their chemical properties are very close, but Zr element is in the IVb group with different crystal structure, electron configuration, bonding radius, et al. This may partially explain different reactivity behavior of $MoSi_2$ and $ZrSi_2$ in the corresponding doped samples.

The transition temperatures for the doped and undoped samples determined by DC susceptibility measurements are shown in Fig. 2. $T_C$ obtained as the onset of magnetic screening for the undoped samples was 36.5 K with a transition width of 2 K. For the doped samples, the $T_C$ decreased with increasing the doping level and the transition became broad. However, the degree of $T_C$ depression by $MoSi_2$ doping is low, indicating that small substitutions occurred in $MgB_2$ crystal. $T_C$ for the 2.5 % and 5 % doped samples showed little difference, which decreased only 0.5 K compared to the undoped ones, while the $T_C$ for the 15 % doped samples dropped about 2 K with a transition width of 6 K. The broad transition in $T_C$ for the 15 % doped samples may be caused by a large amount of impurities such as $MoSi_2$, and this is in agreement with the XRD patterns in Fig. 1.

Figure 3 shows the transport current density at 4.2 K in magnetic fields for the undoped and $MoSi_2$ doped samples annealed at 650 . It is noted that the $MoSi_2$ doping not only enhanced the $J_C$ value in magnetic fields but also improved the $J_C$-B performance. At 2.5 % doping level, the $J_C$ reached 1300 $A/cm^2$ at 4.2K,10 T, more than 4-fold improvement compared to the undoped samples, which have a $J_C$ value about 280 $A/cm^2$. The $J_C$ value of the $MoSi_2$ doped samples decreases with further increasing the doping level, but the $J_C$ field dependence does not change. The possible explanation for the decrease of $J_C$ value with increasing the $MoSi_2$ doping level may be due to the large amount of impurities occurred, which usually lead to weak links at grain boundaries [3], as will be discussed below. On the other hand, the sensitivity of $J_C$ to magnetic field was decreased by the $MoSi_2$ doping, demonstrating an improved flux pinning ability. This means that effective pining centers such as possible segregates or defects were introduced by the $MoSi_2$ doping.

The typical images of the fractured core layers for undoped and doped samples are shown in Fig. 4. SEM results clearly revealed that $MgB_2$ core of the undoped samples was quite porous and loose with some limited melted intergrain regions (see Fig.4a). However, with the addition of 2.5 % $MoSi_2$ (see Fig.4b), the $MgB_2$ core has a quite uniform microstructure with fewer voids. It should be noted that although the 15 % doped samples is still rather dense seeing from the SEM image, there are many large particles with a size of 3~5 μm (see Fig.4 c). These large particles had been identified to be $MoSi_2$ grains by the EDX analysis. The higher the doping level, the more the larger particles scattering within the $MgB_2$ matrix. This is clearly demonstrated by the evident difference in Mo and Si element contents between the 2.5 % and 15 % doped samples from the insets of Fig. 4. Through the EDX analysis at randomly select area on the $MgB_2$ core, we found that $MoSi_2$ ratio is almost identical across the entire sample, suggesting a globally homogeneous phase distribution. As for the 15% doped samples, these largely dispersed $MoSi_2$ particles, which could decrease the superconducting volume of $MgB_2$ tapes and weaken the grain linkage, are proposed to be responsible for the reduction of $J_C$ values.

$J_C$ values vs magnetic filed for the 750 heat-treated samples are plotted in Fig. 5. The $J_C$-B curve of 2.5% doped samples sintered at 650 was also included. It can be seen that again the $J_C$-B properties of the $MoSi_2$ doped samples are improved compared to undoped samples when heated at 750. Like tapes sintered at 650 (see Fig. 3), the best result was achieved for the 2.5 at.% $MoSi_2$ additions. $J_C$ was degraded by further additions although it was still higher than the undoped sample. In addition, when comparing the 2.5% doped-samples heated at different temperatures, we can note that the samples annealed at 750 have slightly better field dependence of $J_C$, implying that a higher sintering temperature leads to more effective flux pinning centers in the doped samples, thus enhancing flux pinning and improving the high-field $J_C$.

Clearly, our data of $MoSi_2$ doped tapes showed much better $J_C$-B performance in the whole range of magnetic fields up to 10 T. Good grain linkage had been thought as one of the reasons for the high $J_C$-B properties in doped tapes. However, as were proved by several groups [9, 14], the grain linkage improvement alone could not explain the lowered field dependence of $J_C$ in doped tapes, since the grain coupling mainly increases the $J_C$ values, hardly changes the field dependence of $J_C$. There must be more effective flux pinning centers in the doped samples than the undoped tapes. It is speculated that the reaction-induced nanoscale impurities or defects induced by the $MoSi_2$ doping could act as the strong pinning centers. It should be noted that the size of $MoSi_2$ particles used was 2-5 μm. In the SiC and C addition, which is significantly effective in increasing $J_C$ in high-field region, nanometer-sized particles were employed [5, 10]. As we know, the surface energy of large particles is much lower than that of nanoscale ones. Thus it is difficult for the micrometer $MoSi_2$ powders to react with Mg or B and to form nanoscale impurities. Therefore, further improvement in $J_C$-B performance is expected upon utilization of finer $MoSi_2$ particles.

## 4. Conclusions

In summary, we have synthesized $MoSi_2$ doped $MgB_2$ tapes by the *in-situ* PIT method. The effect of $MoSi_2$ doping on the microstructures and superconducting properties of $MgB_2$ tapes has been investigated. It is found that the $J_C$ values have been significantly improved by $MoSi_2$ doping. The best result was achieved for the 2.5 at.% $MoSi_2$ additions. $J_C$ was degraded by further additions. Furthermore, the enhanced field dependence of the $MoSi_2$ doped tapes is due to the pinning by possible segregates or defects caused by the $MoSi_2$ doping.

**Acknowledgments**

The authors thank Yulei Jiao, Ling Xiao, Xiaohang Li, S. Awaji and Liye Xiao for their help and useful discussions. This work is partially supported by the National Science Foundation of China under Grant No.50472063 and No.50377040 and National "973" Program (Grant No. 2006CB601004).

# References

1. Nagamatsu J, Nakagawa N, Muranaka T, Zenitani Y and Akimitsu J 2001 *Nature* **410** 63

2. Pachla W, Morawski A, Kovac P, Husek I, Mazur A, LadaT, Diduszko R, Melisek T, Strbık V and Kulczyk M 2006 *Supercond. Sci. Technol.* **19** 1

3. Flukiger R, Suo H L, Musolino N, Beneduce C, Toulemonde P, Lezza P 2003 *Physica C* **385** 286

4. Senkowicz B J, Giencke J E, Patnaik S, Eom C B, Hellstrom E E and Larbalestier D C 2005 *Appl. Phys. Lett.* **86** 202502

5. Ma Yanwei, Zhang Xianping, Nishijima G., Watanabe K, Awaji S and Bai Xuedong, 2006 *Appl. Phys. Lett.* **88** 072502.

6. Berenov A, Serquis A, Liao X Z, Zhu Y T, Peterson D E, Bugoslavsky Y, Yates K A, Blamire M G, Cohen L F, and MacManus-Driscoll J L 2004 *Supercond. Sci. Technol.* **17** 1093

7. Ueda S, Shimoyama J, Yamamoto A, Horii S and Kishio K 2004 *Supercond. Sci. Technol.* **17** 926

8. Wang J, Bugoslavsky Y, Berenov A, Cowey L, Caplin A D, Cohen L F, MacManus-Driscoll J L, Cooley L D, Song X, and Larbalestier D C 2002 *Appl. Phys. Lett.* **81** 2026

9. Ma Yanwei, Zhang Xianping P, Xu A X, Li X H, Xiao L Y, Nishijima G, Awaji S, Watanabe K, Jiao Y L, Xiao L, Bai X D, Wu K H and Wen H H 2006 *Supercond. Sci. Technol.* **19** 133

10. Dou S X, Braccini V, Soltanian S, Klie R, Zhu Y, Li S, Wang X L, and Larbalestier D 2004 *J. Appl. Phy.* **96** 7549

11. Fu B Q, Feng Y, Yan G, Zhao Y, Pradhan A K, Cheng C H, Ji P, Liu X H, Liu C F, Zhou L and Yau K F 2002 *J. Appl .Phys.* **92** 73

12. Ma Yanwei, Kumakura H, Matsumoto A and Togano K 2003 *Appl.Phys.Lett.* **83** 1181

13. Matsumoto A, Kumakura H, Kitaguchi H and Hatakeyama H 2004 *Supercond.Sci.Technol.* **17** S319

14. Jiang C H, Nakane T and Kumakura H 2005 *Supercond. Sci. Technol.* **18** 902

15. Li S , White T, Sun C Q, Fu Y Q, Plevert J and Lauren K 2004 *J. Phys. Chem. B* **108** 16415

# Captions

Figure 1  XRD patterns of in-situ processed undoped and all MoSi$_2$ doped tapes heated at 650 ℃ for 1 h. The peaks of MgB$_2$ indexed, while the peaks of MgO and MoSi$_2$ are marked by asterisks and circles, respectively.

Figure 2  The temperature dependence of the DC magnetic susceptibility curves of the MoSi$_2$ doped and undoped tapes.

Figure 3  $J_C$-B properties of Fe-sheathed undoped and MoSi$_2$ doped tapes heated at 650 ℃ for 1 h.

Figure 4  SEM images of the undoped (a), 2.5% (b), and 15% (c) MoSi$_2$ doped samples after peeling off the Fe sheath.

Figure 5  $J_C$-B properties of Fe-sheathed undoped and MoSi$_2$ doped tapes heated at 750 ℃ for 1 h. $J_C$-B curve of 2.5% doped samples heated at 650 ℃ was also included.

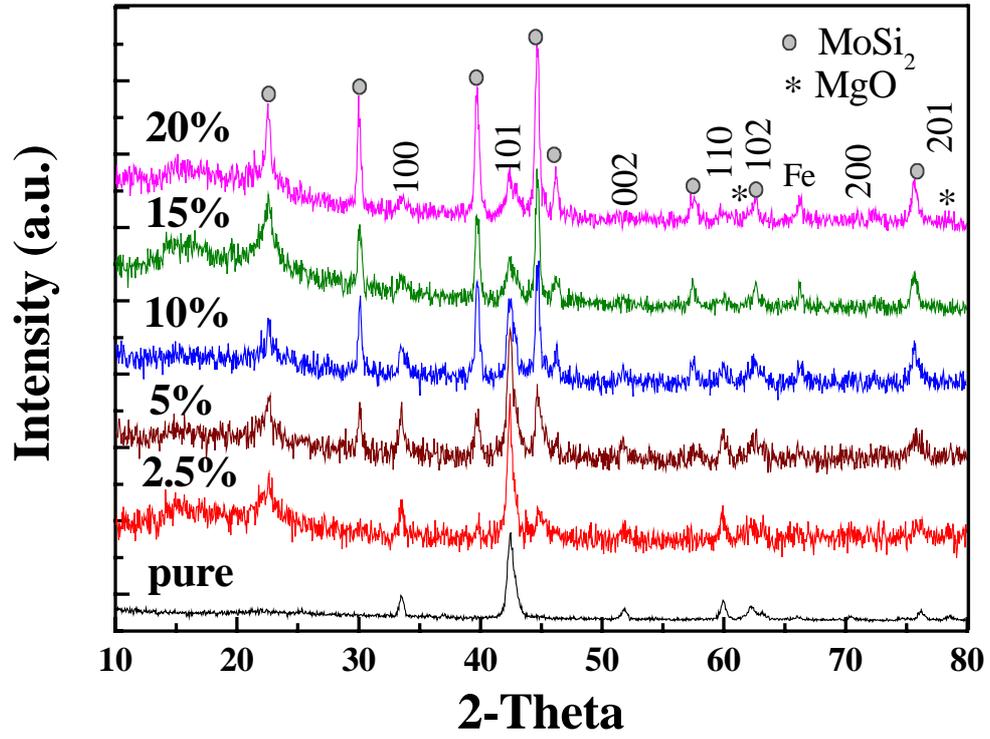

Fig.1 Zhang et al.

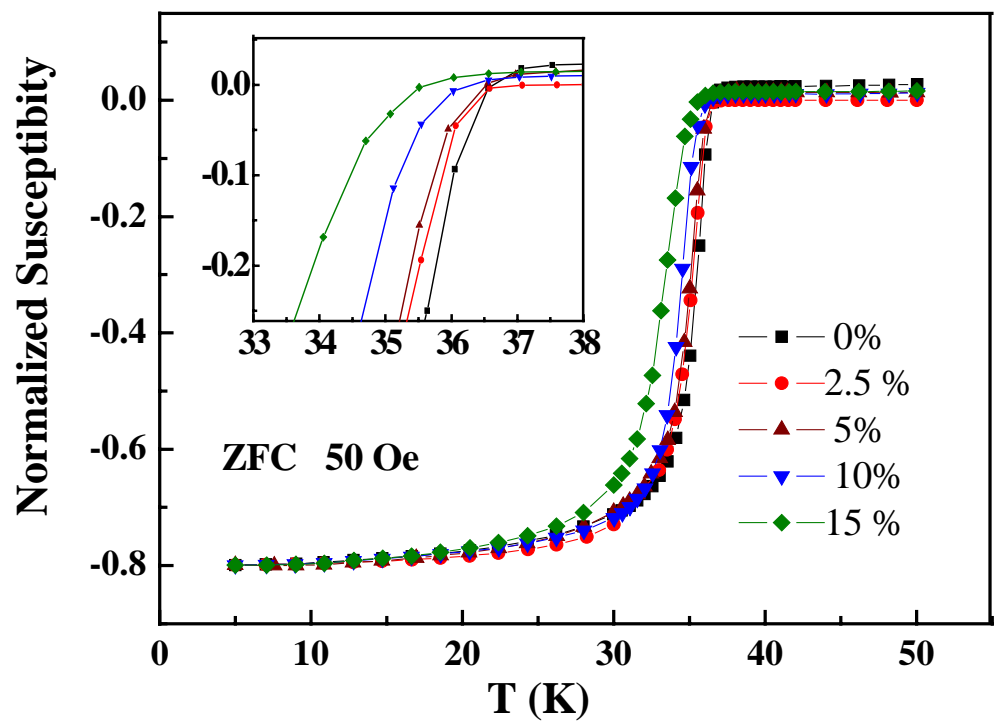

Fig 2 Zhang et al.

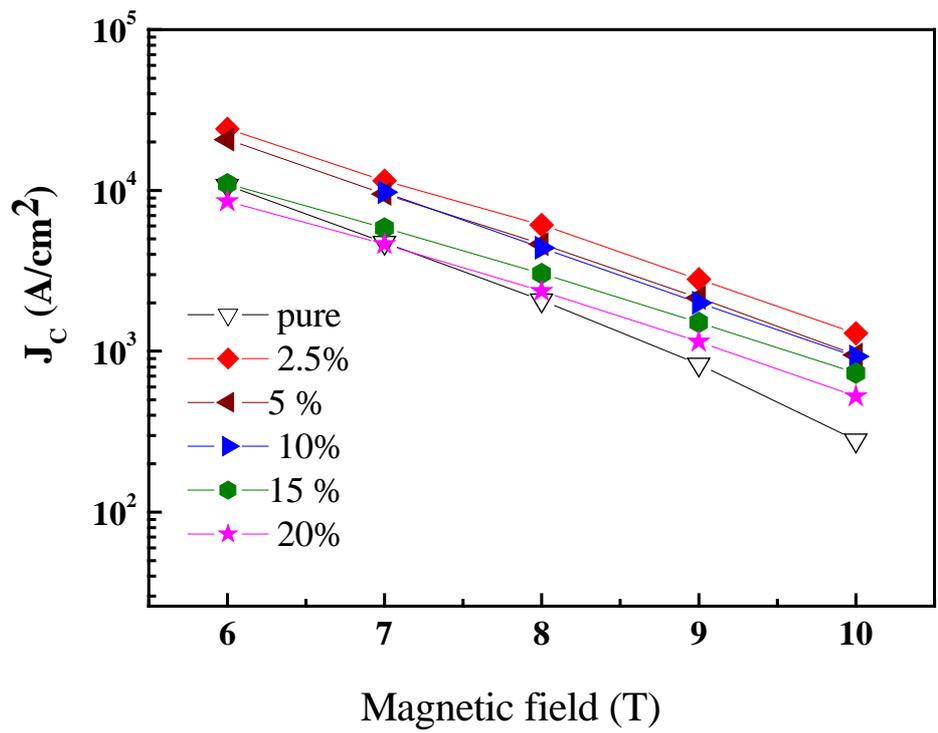

Fig 3 Zhang et al.

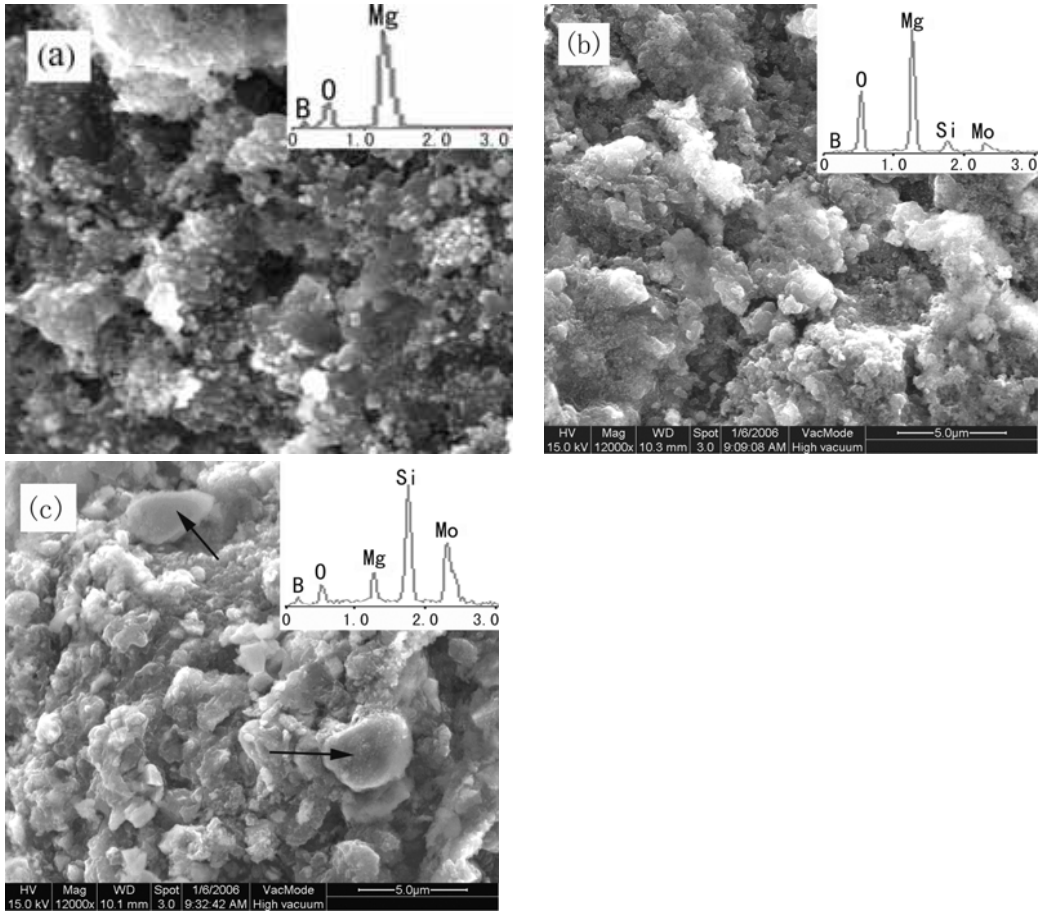

Fig 4 Zhang et al.

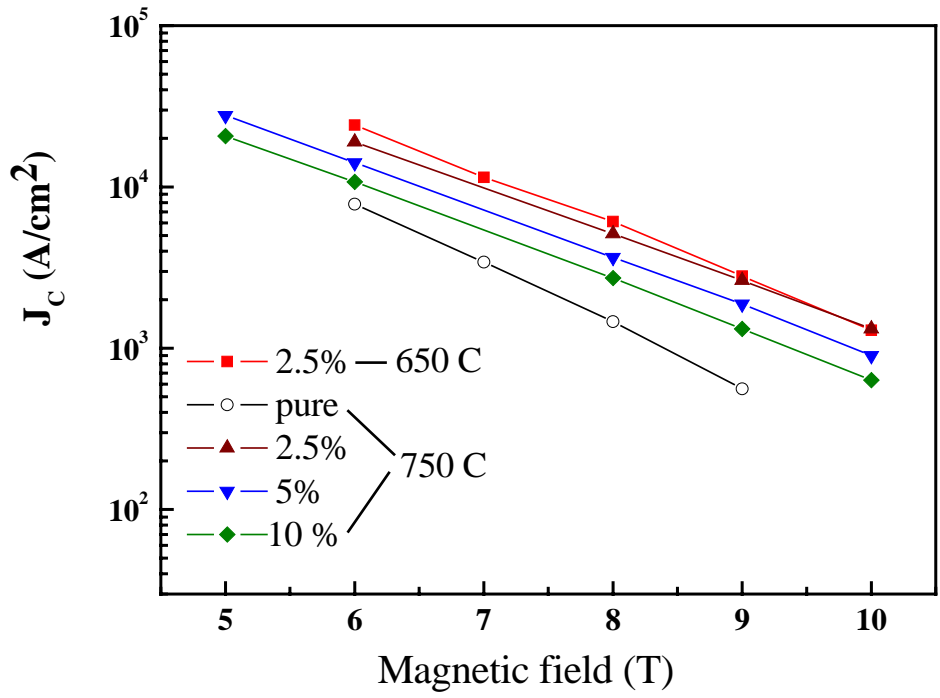

Fig 5 Zhang et al.